\documentstyle[12pt,equation]{article}
\begin{document}
\newcommand{\be}{\begin{equation}}
\newcommand{\ee}{\end{equation}}
\newcommand{\een}{\end{subequations}}
\newcommand{\ben}{\begin{subequations}}
\newcommand{\beq}{\begin{eqalignno}}
\newcommand{\eeq}{\end{eqalignno}}
\newcommand{\ver}{\mbox{$\gamma \bar{q} q$}}
\newcommand{\xg}{\mbox{$x_{\gamma}$}}
\renewcommand{\thefootnote}{\fnsymbol{footnote} }
\pagestyle{empty}
\setcounter{page}{1}
\begin{flushright}
MAD/PH/797 \\
October 1993\\
\end{flushright}

\vspace{1.5cm}
\begin{center}
{\Large \bf Initial State Showering in Resolved Photon
Interactions}\footnote{Talk held at the {\it 23rd International Symposium on
Multiparticle Dynamics}, Aspen, Colo., Sep. 1993}\\
\vspace{5mm}
Manuel Drees\footnote{Heisenberg fellow}\\
{\em Physics Department, University of Wisconsin, 1150 University Ave,
Madison, WI 53706, USA}
\end{center}

\vspace{2cm}
\begin{abstract}
\noindent
After a brief review of recent data that confirm qualitative and
quantitative predictions for resolved photon processes, a possible problem
with the implementation of initial state showering in existing Monte Carlo
event generators is pointed out. It is argued that this is responsible for
the rather poor description of the jet rapidity distribution measured by the
H1 collaboration at HERA.
\end{abstract}
\end{document}